\documentclass[reprint,amsmath,amssymb,aps,prl,twocolumn,longbibliography]{revtex4-2}

\usepackage{float}
\usepackage{braket}
\usepackage{graphicx}
\usepackage{dcolumn}
\usepackage{bm}
\usepackage[colorlinks=true, urlcolor=blue, linkcolor=blue, citecolor=blue]{hyperref}
\usepackage[utf8x]{inputenc}
\usepackage{xcolor} 
\usepackage[normalem]{ulem}

\newcommand{\affcua}{MIT-Harvard Center for Ultracold Atoms, Research Laboratory of Electronics, and Department of Physics, Massachusetts Institute of Technology, Cambridge, Massachusetts 02139, USA}

\begin{document}

\title{Measuring pair correlations in Bose and Fermi gases via atom-resolved microscopy}

\author{Ruixiao Yao}
\author{Sungjae Chi}
\author{Mingxuan Wang}
\author{Richard J. Fletcher}
\author{Martin Zwierlein}

\affiliation{\affcua}

\date{\today}

\begin{abstract}
We demonstrate atom-resolved detection of itinerant bosonic $^{23}$Na and fermionic $^6$Li quantum gases, enabling the direct in situ measurement of interparticle correlations. In contrast to prior work on lattice-trapped gases, here we realize microscopy of quantum gases in the continuum. We reveal Bose-Einstein condensation with single-atom resolution, measure the enhancement of two-particle $g^{(2)}$ correlations of thermal bosons, and observe the suppression of $g^{(2)}$ for fermions; the Fermi or exchange hole. For strongly interacting Fermi gases confined to two dimensions, we directly observe non-local fermion pairs in the BEC-BCS crossover. We obtain the pairing gap, the pair size, and the short-range contact directly from the pair correlations. In situ thermometry is enabled via the fluctuation-dissipation theorem. Our technique opens the door to the atom-resolved study of strongly correlated quantum gases of bosons, fermions, and their mixtures.
\end{abstract}

\maketitle

Quantum many-body physics is at its heart the study of interparticle correlations. These can be of purely quantum statistical origin, as in the case of the ideal Bose and Fermi gases, or they can be brought about by interactions between particles. The interplay of quantum statistics and strong correlations renders many problems of interest highly challenging to solve theoretically. The idea of quantum simulation is to employ ultracold quantum gases of atoms or molecules to realize model systems for such strongly interacting matter, from high-temperature superfluids~\cite{Giorginireview, Varennanotes, Zwerger2012} to quantum magnets~\cite{Bloch2008} and topological systems~\cite{Cooper2019}.

A breakthrough for experiments on lattice-trapped atomic gases, realizing Hubbard models or spin physics, was the implementation of single-atom, single-lattice site resolved imaging for bosons~\cite{Bakr2009,Sherson2010} and fermions~\cite{Cheuk2015,Haller2015,Parsons2015,Omran2015,Edge2015,Gross2017}. These quantum gas microscopes have enabled the direct measurement of particle correlations, revealing anti-ferromagnetic correlations~\cite{Parsons2016,Cheuk2016,Boll2016} and fermion pairing~\cite{Hartke2023} in Hubbard systems.

 \begin{figure}
 \includegraphics[width=1\linewidth]{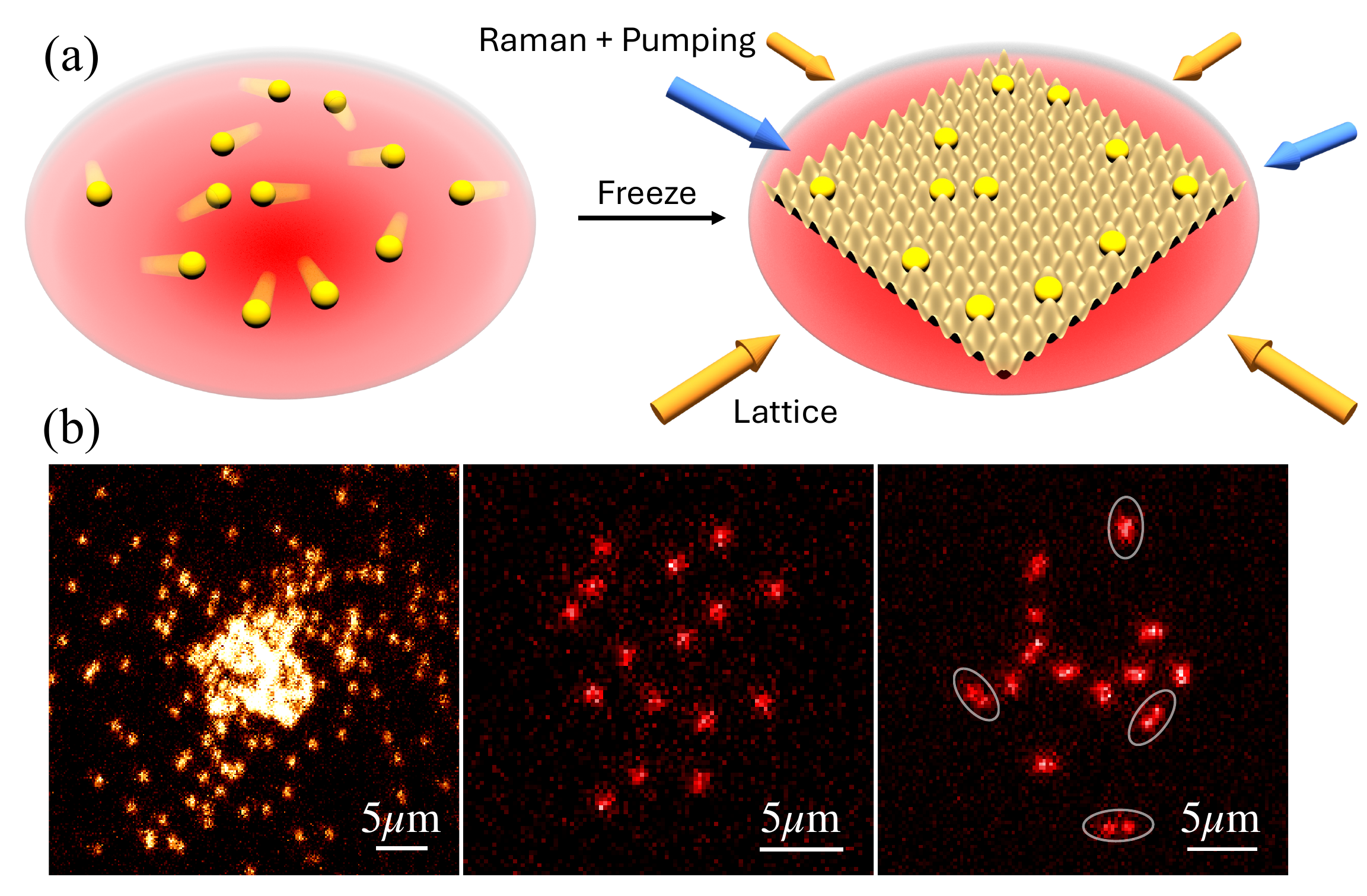}
   \caption{
   Atom-resolved microscopy of quantum gases in the continuum. (a) Itinerant atoms in an atom trap (red) are suddenly frozen in place via an applied optical lattice and imaged via Raman sideband cooling~\cite{Cheuk2015}.
   (b) Microscope images of bosonic $^{23}$Na forming a Bose-Einstein condensate (left), of a single spin state in a weakly interacting $^{6}$Li Fermi mixture (middle), and of both spin states of a strongly interacting Fermi mixture, directly revealing pair formation (right).
   } 
 \label{fig1:microscope}
\end{figure}

However, paradigmatic systems of many-body physics, e.g. weakly interacting Bose gases \cite{Hadzibabic2011}, or strongly interacting Fermi gases in the BEC-BCS crossover \cite{Giorginireview, Varennanotes, Zwerger2012}, exist in the continuum, without an underlying lattice potential. Their study thus far involved more coarse-grained probes, principally absorption imaging. Correlations in momentum space were observed with single-atom detection using time-of-flight expansion techniques, from the enhancement and suppression of two-particle correlations in Bose and Fermi gases~\cite{Jeltes2007}, to noise correlations~\cite{Folling2005,Greiner2005} and recently momentum correlations of atom pairs in strongly interacting few-fermion systems~\cite{Holten2022}.

Here we demonstrate real-space, atom-resolved microscopy of quantum gases in the continuum, and apply the technique to two paradigmatic systems, the weakly interacting two-dimensional (2D) Bose gas and the strongly interacting 2D Fermi gas in the BEC-BCS crossover. Imaging quantum gases in situ at the resolution of single atoms realizes the ultimate depth of information one may obtain in real space. Not only may one obtain thermodynamic quantities such as density, compressibility and pressure~\cite{Ku2012,Desbuquois2014}, which were previously accessible with coarse-graining probes, but also interparticle correlations of in principle arbitrary order are accessible. The technique developed here is general and can be applied to other continuum systems of interest, from spin-imbalanced Fermi gases~\cite{Zwierlein2016a} putatively hosting the Fulde-Ferrell-Larkin-Ovchinnikov (FFLO) phase of finite-momentum Cooper pairs ~\cite{Fulde1964,Larkin1964,Radzihovsky2010}, to Bose-Fermi mixtures with their intricate phase diagram~\cite{Ludwig2011,Bertaina2013}, dipolar atomic~\cite{Chomaz2023} and molecular~\cite{Bigagli2024} quantum gases hosting supersolids and topological superfluids~\cite{Cooper2009}, and to the wide array of impurity problems such as Bose and Fermi polarons~\cite{Massignan2025}.

Here we focus on two-particle correlations, revealing the enhanced $g^{(2)}$ correlations of a thermal Bose gas, the suppressed correlations of a Fermi gas, known as the Fermi hole~\cite{Wigner1934,Cheuk2016,Hartke2020,Paulihole}, and the formation of non-local fermion pairs in the BEC-BCS crossover. We directly obtain the equal spin and density-density correlations as a function of interaction strength, giving access also to interspin correlations and, importantly, the contact characterizing short-range correlations~\cite{Tan2008, Werner2012, Bertaina2011}.

The idea of the measurement technique is sketched in Fig.~\ref{fig1:microscope}(a). A quantum gas of $^{23}$Na bosons or $^6$Li fermions explores the continuum of an atom trap. At the time of the measurement, the atoms' position is suddenly frozen by ramping up a pinning lattice, capturing atoms in the wells of the optical lattice potential. A light sheet ensures tight transverse confinement. Fluorescence from atoms is collected via Raman sideband cooling~\cite{Cheuk2015, Parsons2015,Brown2017}. The concept was recently demonstrated by the Yefsah group in the study of expanding matter waves of single atoms, and regimes of high-fidelity read-out were characterized~\cite{Verstraten2024}.
The method most naturally lends itself to the study of quasi-2D gases that we focus on here, but can in principle be extended to 3D gases.
Our trap is formed by a single tightly-focused 1070 nm laser beam at vertical waist $w = 4\,\mu$m, leading to near-circularly symmetric confinement in the horizontal plane with trapping frequencies for $^6$Li $\nu_{x,y,z} = (110(7),94(1),3095(35))\rm\,Hz$. The pinning lattice is derived from a 1064 nm laser and set up in a retro-reflected bow-tie configuration~\cite{Brown2017}, enhancing the trap depth in each well. The resulting square lattice has a spacing of $752 {\rm\,nm}$.
Atom pairs at short distance may end up in the same lattice well, and their density information is lost due to photoassociation. Techniques to protect pairs against such parity-projecting loss have been developed, e.g. via bilayer imaging~\cite{Hartke2023}. To obtain a faithful measurement of atom positions, largely unaffected by the finite resolution offered by the pinning lattice spacing, we work with dilute gases with typical interparticle spacings of $n^{-1/2} > 3\,{\rm \mu m}$.

Fig.~\ref{fig1:microscope}(b) (left) shows the method applied to a Bose-Einstein condensate (BEC) of $^{23}$Na, containing about 100 atoms. Raman sideband cooling for $^{23}$Na has been implemented in optical tweezers~\cite{Yu2018}, and the method is adapted here for a continuum quantum gas microscope~\cite{Cheuk2015}.
The same apparatus also produces fermionic $^6$Li atoms in arbitrary mixtures of two hyperfine states, realizing 2D Fermi gases with tunable interparticle interactions. Fig.~\ref{fig1:microscope}(b) (middle and right) show typical images of Fermi gases with weak and strongly attractive interactions. Already in individual images, the tendency of fermions to anti-bunch and of strongly attractively interacting fermions to reorganize in closely-spaced pairs is apparent.

 \begin{figure}
 \includegraphics[width=1\linewidth]{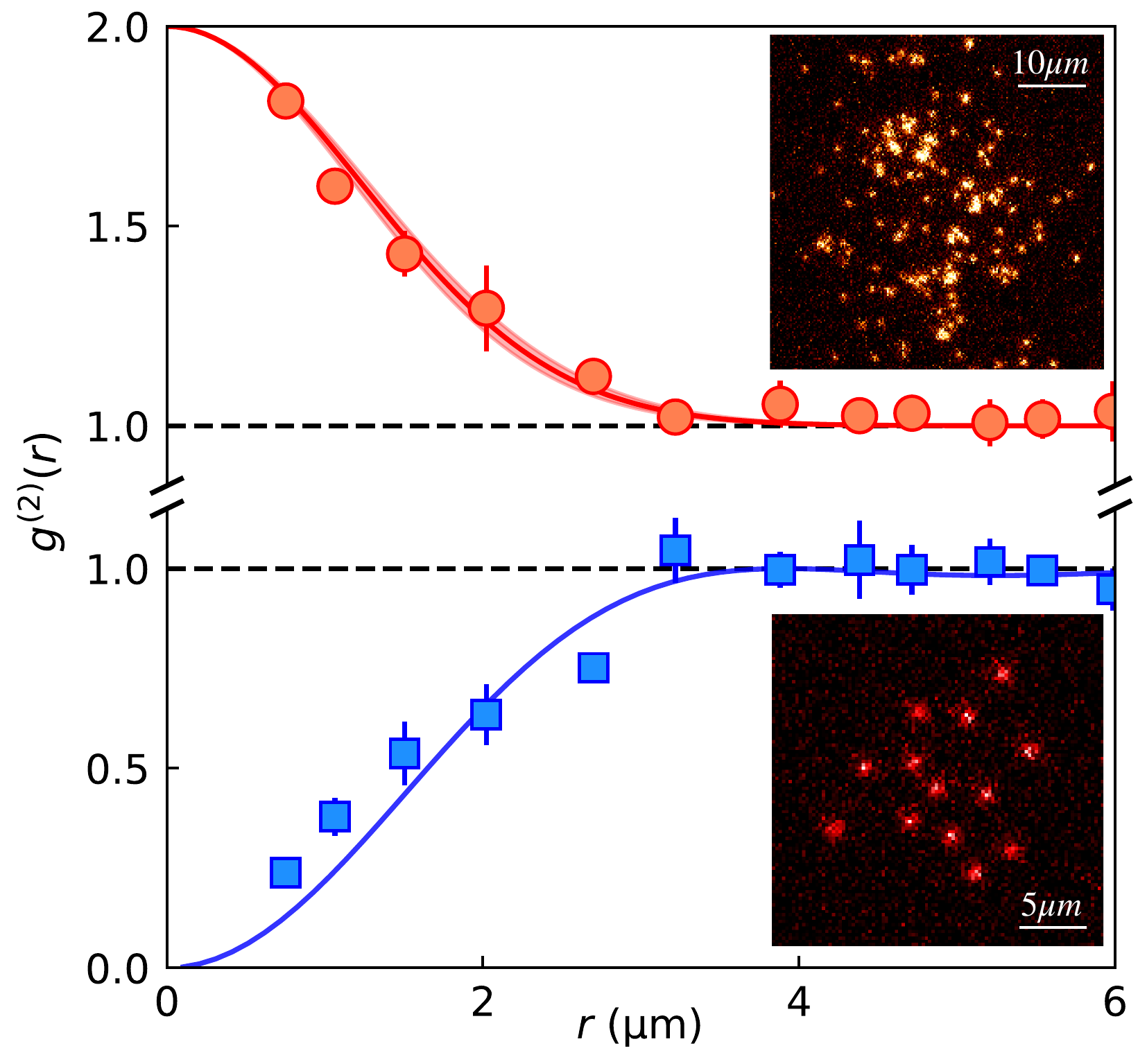}
   \caption{
    Pair correlation function of a thermal Bose gas (top) and a non-interacting Fermi gas (bottom).
    The red curve (top) is a fit giving a thermal de Broglie wavelength of $\lambda_{\rm dB}{=}4.4\,\rm \mu m$ and a temperature $T{=}6.9(3)$\,nK. The blue curve (bottom) is the $T{=}0$ pair correlation for a 2D non-interacting Fermi gas at our interparticle spacing $n^{-1/2}{=}3.6\,\rm \mu m$, without free parameters. The insets show exemplary microscope images for the Bose and Fermi gas. Black dashed lines indicate $g^{(2)}{=}1$. 
   } 
 \label{fig2:BoseFermicorrelations}
\end{figure}

As a first application of the atom-resolved imaging, we measure in Fig.~\ref{fig2:BoseFermicorrelations} the two-particle correlations of a thermal Bose gas and of a deeply degenerate Fermi gas.
A famous consequence of quantum statistics is that thermal bosons display the tendency to bunch~\cite{Naraschewski1999}, while fermions display anti-bunching: The probability to find two bosons near each other is enhanced above mere chance, while for fermions it is reduced, a phenomenon known as the Fermi hole~\cite{Wigner1934,Paulihole}. In the context of quantum gases, the Fermi hole was observed for lattice-trapped fermions in~\cite{Cheuk2016, Hartke2020}. For continuum quantum gases, the correlation peak for bosons or hole for fermions has not been observed in situ before. The complementary phenomenon in momentum space was observed for bosons and fermions in~\cite{Jeltes2007}.

The size of the correlation peak or hole is given by the de Broglie wavelength of the particles. For fermions, the de Broglie wavelength cannot exceed the interparticle spacing. For bosons, one is restricted to temperatures above $T_c$ or, in a trapped gas, to regions outside the BEC, as the latter is second-order coherent ($g^{{2}}{=}1$). So for both statistics, the size of the correlation peak or hole is limited by the interparticle spacing. To be observable under the microscope, we thus need to work with interparticle distances larger than the spacing of the pinning lattice.
For the weakly interacting Bose gas of $^{23}$Na, the low density presents a challenge for thermalization. We circumvent this by starting with a BEC, applying controlled heating via a parametric drive for 360\,ms, and equilibrating for one second to obtain a low-density thermal cloud close to equilibrium.
The Fermi gas, on the other hand, is produced near a Feshbach resonance, with excellent thermalization rates. For the data in the figure, the final interaction strength is $\eta{=}\log(k_{\rm F} a_{\rm 2D}){=}4.2$, where $a_{\rm 2D}$ is the 2D scattering length~\cite{Petrov2001,Bertaina2011}, related to the 3D scattering length $a_{\textrm{3D}}$ via $a_{\textrm{2D}}{=}2.093\, a_z \exp{(-\sqrt{\frac{\pi}{2}}a_z/a_{\textrm{3D}})}$, where $a_z = \sqrt{\hbar/m\omega_z}$ is the transverse harmonic oscillator length.
Sizeable correlation peaks and holes are observed in the $g^{(2)}$ correlation function for bosons and fermions (Fig.~\ref{fig2:BoseFermicorrelations}).
The $g^{(2)}$ function for ideal bosons ($\varepsilon{=}{+1}$) and fermions ($\varepsilon{=}{-1}$) is given by $ g^{(2)}(r){=}1{+}\varepsilon \frac{1}{n^2}\left|\sum_k n_k e^{i \vec{k}\cdot\vec{r}}\right|^2 $
where $n_k$ is the momentum distribution, $\sum_k{=}\int\frac{{\rm d}^2k}{(2\pi)^2}$ and $n{=}\sum_k n_k$ the density.
For the Bose gas data, a good fit is achieved for the approximation $g^{(2)}(r) = 1 + \exp(-2\pi r^2/\lambda_{\rm dB}^2)$ valid in the non-degenerate regime~\cite{Naraschewski1999}. We obtain a thermal de Broglie wavelength $\lambda_{\rm dB} = 4.4\,\rm\mu m$, corresponding to a temperature of $T = 6.9(3)\rm\,nK$.
A full study of two-particle correlations in the degenerate interacting 2D Bose gas~\cite{Hadzibabic2011} is an important problem for future studies, in particular as the gas crosses over into a Berezinskii-Kosterlitz-Thouless superfluid, where $g^{(2)}(r)$ is expected to display algebraic decay. In the presence of a trap, condensate formation should restore second-order coherence, but strong interatomic repulsion will modify $g^{(2)}(r)$~\cite{Ceperley1995}.

The weakly interacting Fermi gas data closely match the theoretical form $g^{(2)}(r){=}1{-}\left|\frac{2}{k_{\rm F} r}J_1(k_{\rm F} r)\right|^2$ for a non-interacting Fermi gas of density $n_\uparrow$ in two dimensions at zero temperature, with measured $k_{\rm F}{=}\sqrt{4\pi n_\uparrow}{=}0.98 \,\mu m^{-1}$. 
The reduced $g^{(2)}$ probability immediately implies sub-Poissonian fluctuations of the Fermi gas~\cite{Muller2010,Sanner2010}. Indeed, we observe $\Delta N^2/N = 0.46(5)$ for this dataset. Fluctuation-dissipation thermometry~\cite{Hartke2020,suppmat} gives a temperature $T{=}6.1(2)\,\rm nK$, and with $E_F = k_B\cdot 39\,\rm nK$ a reduced temperature of $T/T_F{=}0.16$.

 \begin{figure}
 \includegraphics[width=1\linewidth]{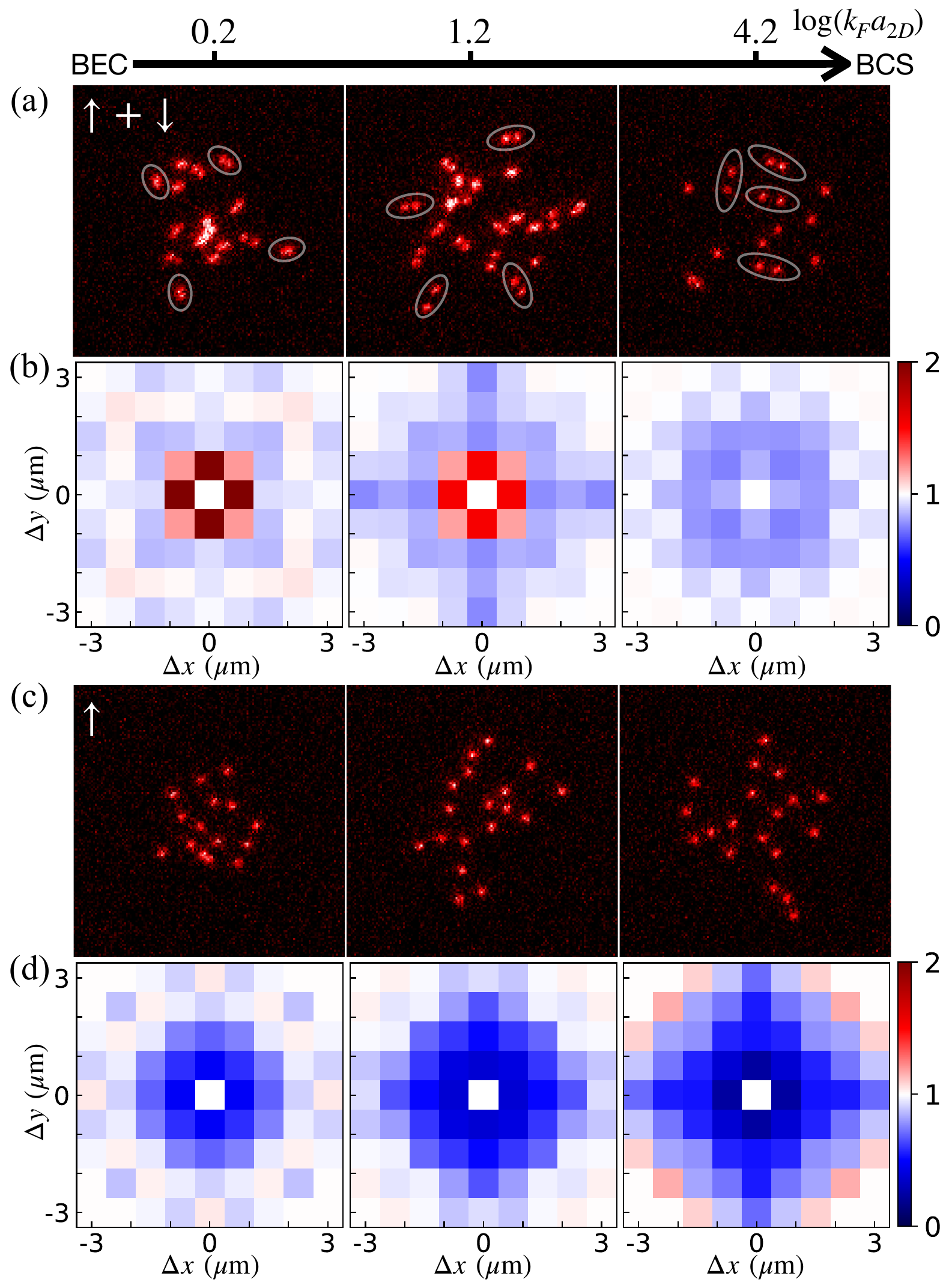}
   \caption{
   Pair correlations of the 2D strongly interacting Fermi gas in the BEC-BCS crossover.
   (a) Fermi gas microscope images of both spin states from the BEC to the BCS regime ($\eta=\log(k_{\rm F} a_{\rm 2D}){=}0.2$, 1.2, and 4.2 from left to right). 
   The thin ellipses show closely spaced pairs of fermions, as expected in the BEC-BCS crossover.
   (b) The density-density correlation map $g_{nn}^{(2)}(\vec{r})$, showing how the pair size increases from the BEC to the BCS regime.
   (c) Microscope images with one spin component removed.
   (d) The $\uparrow\uparrow$ correlation map for a single spin component. The Fermi hole grows towards the BCS limit.
   } 
 \label{fig3:BECBCSimages}
\end{figure}

We now turn to the study of strongly interacting Fermi gases in two dimensions, in the crossover from Bose-Einstein condensation of tightly bound molecules to BCS superfluidity of long-range Cooper pairs.
Following studies in three dimensions~\cite{Zwerger2012, Giorginireview, Varennanotes}, a wealth of experimental results has already been gathered for the 2D case, from the study of the pairing energy~\cite{Feld2011,Sommer2012}, the equation of state~\cite{Makhalov2014,Vale2016EoS, Enss2016EoS}, radiofrequency spectra revealing the contact~\cite{ZwergerClockShift2012} to evidence for condensation~\cite{Ries2015}.
Here we directly observe the equal spin and density-density correlation function from our atom-resolved microscope images, yielding important microscopic information about this strongly correlated Fermi system.

In Fig.~\ref{fig3:BECBCSimages}(a), we show images of the total density of the spin-balanced mixture. Fermion pairing is apparent in the pairwise clustering of atoms. The pair size increases from the BEC to the BCS limit of the crossover, as expected.
The corresponding density-density correlation function as a function of distance between pairs is shown in Fig.~\ref{fig3:BECBCSimages}(b). We observe bunching at short-range due to pairing, while at distances on the order of the interparticle spacing, the fermionic anti-bunching due to the Fermi hole between like spins dominates.
The equal spin ($\uparrow\uparrow$) correlation function can be obtained by removing one spin state ($\downarrow$) after pinning using resonant light (applied at $B= 843\,\rm G$). This technique was previously used in the study of lattice gases~\cite{Brown2017} and has been shown to preserve spin up atoms even if they were co-trapped with a down spin in the same lattice site before the light pulse. The images shown in Fig.~\ref{fig3:BECBCSimages}(c) thus faithfully show all $\uparrow$ atoms, even those that were part of a short-range pair.
These single-spin data clearly reveal the Fermi hole persisting for all interaction strengths (Fig.~\ref{fig3:BECBCSimages}(d)). It is also apparent that the size of the exchange hole shrinks towards the BEC side. This is expected, as pairing will broaden the momentum distribution of fermions and thus shrink the region in real space affected by Pauli exclusion.

 \begin{figure*}
 \includegraphics[width=0.8\linewidth]{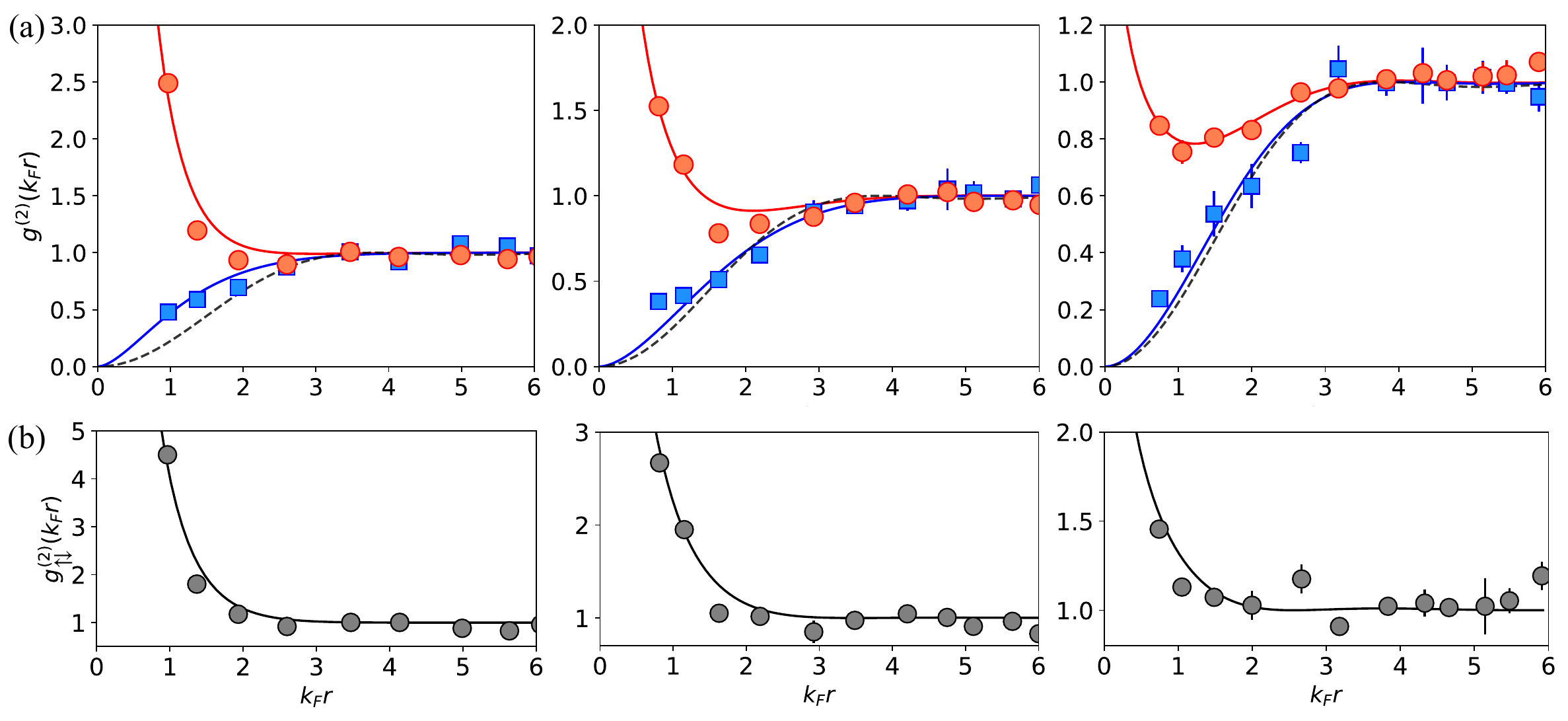}
   \caption{
   a) Pair correlation functions for total density $g_{nn}^{(2)}$ (red circle) and equal spin $g_{\uparrow\uparrow}^{(2)}$ (blue square) in the BEC-BCS crossover. From left to right $\eta=\log(k_{\rm F} a_{\rm 2D})=0.2$, $1.2$ and $4.2$. The red and blue solid curves are fits to Eqs.~\ref{eq:correlations}. Black dashed curve: correlation function for ideal Fermi gas at $T=0$.
   b) Derived unequal spin correlation function $g_{\uparrow\downarrow}^{(2)} = 2 g_{nn}^{(2)} - g_{\uparrow\uparrow}^{(2)}$.
   } 
 \label{fig4:correlators}
\end{figure*}

In Fig.~\ref{fig4:correlators}(a), we display the measured density-density correlation function $g_{nn}^{(2)}(\vec{r}{-}\vec{r}\,') = \frac{\left<n(\vec{r})n(\vec{r}\,'\right>}{\left<n(\vec{r})\right>\left<n(\vec{r}\,')\right>}$ and equal spin correlation function $g_{\uparrow\uparrow}^{(2)}(\vec{r}{-}\vec{r}\,')=\frac{\left<n_\uparrow(\vec{r})n_\uparrow(\vec{r}\,'\right>}{\left<n_\uparrow(\vec{r})\right>\left<n_\uparrow(\vec{r}\,')\right>}$ versus distance, normalized by the inverse Fermi wavevector $k_{\rm F}^{-1}$, where $k_{\rm F}{=}\sqrt{2\pi n}$, with $n$ the total density. The Fermi hole in $g_{\uparrow\uparrow}^{(2)}$ is visible throughout, and for strong interactions, the tendency to form closely spaced pairs is already evident in $g_{nn}^{(2)}$. In Fig.~\ref{fig4:correlators}(b) we show the deduced unequal spin correlation function $g_{\uparrow\downarrow}^{(2)} = 2 g_{nn}^{(2)} - g_{\uparrow\uparrow}^{(2)}$ which makes short-range pair correlations evident even for $\eta = 4.2$.

To further interpret these correlation functions, we first recall their derivation within the mean-field BEC-BCS crossover theory~\cite{Miyake1983,Randeria1989}. One introduces quasi-particle amplitudes $u_k$ and $v_k$, finding $(u_k^2,v_k^2){=}\frac{1}{2}\left(1 \pm \frac{\xi_k}{E_k}\right)$ with $\xi_k{=}\frac{\hbar^2 k^2}{2m}{-}\mu$ and a quasi-particle dispersion relation $E_k{=}\sqrt{\xi_k^2 + \Delta^2}$. The number and gap equations yield $\Delta{=}\sqrt{2E_F E_B}$ and $\mu{=}E_F{-}E_B/2$ with $E_B{=}\frac{\hbar^2}{m a_{\rm 2D}^{*2}}$ the two-body binding energy and $a_{\rm 2D}^*{=}a_{\rm 2D} e^\gamma/2$ (with $\gamma{=} 0.577\dots$) related to the 2D scattering length $a_{\rm 2D}$.
Evaluating the correlation functions yields $g_{\uparrow\downarrow}^{(2)}(\vec{r}){=}1{+}\frac{1}{n_\uparrow n_\downarrow}\left|\sum_k u_k v_k e^{i \vec{k}\cdot \vec{r}}\right|^2$ and $g_{\uparrow\uparrow}^{(2)}(\vec{r}){=}1{-}\frac{1}{n_\uparrow^2}\left|\sum_k v_k^2 e^{i \vec{k}\cdot\vec{r}}\right|^2$ with $n_\sigma{=}\sum_k v_k^2$. The sums can be done analytically~\cite{Obeso2022} and give:
\begin{eqnarray}
    g_{\uparrow\downarrow}^{(2)}(r) &=& 1 + 4 c  \left|J_0(k r)K_0(r/b)\right|^2\\
    g_{\uparrow\uparrow}^{(2)}(r) &=& 1 - 4 c\left|J_1(k r)K_1(r/b)\right|^2 
\label{eq:correlations}
\end{eqnarray}
which together yield $g_{nn}^{(2)}{=}\frac{1}{2} g_{\uparrow\uparrow}^{(2)}{+}\frac{1}{2} g_{\uparrow\downarrow}^{(2)}$.
Here, $c{=}C/k_{\rm F}^4$ is the dimensionless short-range contact~\cite{Tan2008, Werner2012}, and $J_i$, $K_i$ are Bessel functions.
These forms ensure that $g_{\uparrow\uparrow}^{(2)}(0){=}0$ as required for equal-spin fermions and $\lim_{r\rightarrow 0} g_{\uparrow\downarrow}^{(2)}(r){=}4c \log(r/a_{\rm 2D})^2$, the short-range behavior dictated by two-body physics.
Within mean-field, the parameters $k{=}k_{\rm F}$, $b{=}a_{\rm 2D}^*$, and the dimensionless contact $c{=}1/(k_{\rm F} a_{\rm 2D}^*)^2$ is solely due to two-body pairing.

Inspired by the mean-field result, we use the same functional form as Eqs.~\ref{eq:correlations} to fit our correlators, with the contact $c$ and an effective pair size $b$ as the fit parameters, and keeping the constraint $c{=}1/(kb)^2$ to still ensure the correct limiting behaviors as $r{\rightarrow}0$~\footnote{For our correlation function $\lim_{r\rightarrow 0} g_{\uparrow\downarrow}^{(2)}(r) {=} 4c \log(r/\tilde{a}_{\rm 2D})^2$, which has the correct $\propto \log(r)^2$ behavior but $\tilde{a}_{\rm 2D} \ne a_{\rm 2D}$ in general, leading to a constant difference to the two-body limit that becomes negligible as $r\rightarrow 0$.}.
The resulting fits reported in Fig.~\ref{fig4:correlators} are excellent for all interaction strengths explored.

 \begin{figure}
 \includegraphics[width=0.98\linewidth]{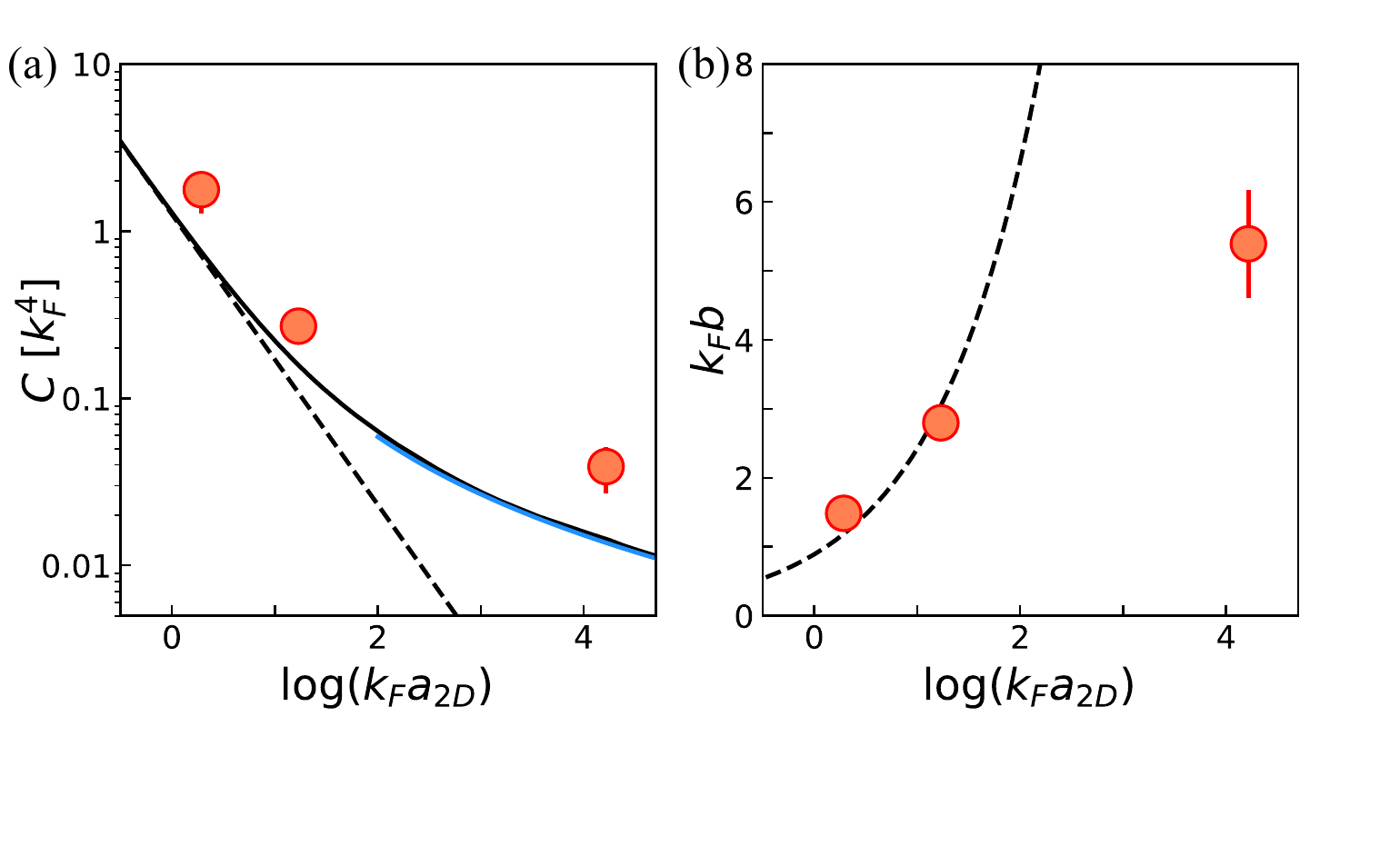}
   \caption{
    Characterization of pairing in the BEC-BCS crossover. a) Contact, b) effective pair size as obtained from fits to correlation functions Eqs.~\ref{eq:correlations}.
    In a), black solid line: Monte Carlo result~\cite{Shi2015}. black dashed line: mean field result. Blue solid line: Fermi liquid contact~\cite{Shi2015}. In b) black solid line: mean-field result $k_{\rm F} a_{\rm 2D}e^{\gamma}/2$.
   } 
 \label{fig5:results}
\end{figure}

From the short-range behavior of $g_{nn}^{(2)}(r)$ we thus directly obtain the contact (see Fig.~\ref{fig5:results}(a)), a crucial quantity connecting the high-momentum tails of the momentum distribution~\cite{Tan2008, Werner2012}, the slope of the energy per particle with respect to interaction strength, the rf clock shift~\cite{Baym2007,ZwergerClockShift2012} and high-frequency tails of rf spectra~\cite{Lucas2010, Mukherjee2019}. The contact of the 2D Fermi gas has previously been measured in the Fermi liquid regime via rf spectroscopy~\cite{Giamarchi2012}. Our data clearly disagrees with the mean-field result, but is in reasonable agreement with Monte-Carlo calculations~\cite{Bertaina2011,Shi2015}. The slight overestimate may be due to the assumption of the BCS form of correlations Eqs.~\ref{eq:correlations} and the fact that the shortest distances probed are not much smaller than the interparticle spacing.

We note that observing a non-zero contact in the BCS regime of large $\eta$ does not imply fermion pairs to be condensed, i.e. superfluidity. Even a normal 2D Fermi liquid will display pair correlations and a contact $C \sim n_\uparrow n_\downarrow/\eta^2$~\cite{Shi2015}, which Monte-Carlo studies show to be largely insensitive to temperature even up to temperatures $T \sim T_F$~\cite{He2022}~\footnote{The BCS Ansatz, which ignores interactions in the normal state, incorrectly associates the presence of a contact to a non-zero pairing gap $c \sim \Delta^2$~\cite{Zwierlein2016a}}.
The parameter $b$ captures an effective pair size and is reported in Fig.~\ref{fig5:results}(b). As the contact is larger than predicted by mean-field, the pair size is correspondingly smaller than the mean-field result, the size of the two-body bound state $a_{\rm 2D}^*$. For the weakest attraction at $\eta{=}4.2$ we obtain a pair size significantly larger than the interparticle spacing, as expected in the BCS regime. However, we caution that in this regime the gas may not be superfluid at our temperatures and the parameter $b$ may rather capture a typical range of Fermi liquid pair correlations.
We note that modifications due to the confining trap should be small, as the Fermi energy is an order of magnitude larger than the harmonic energy, $E_F \sim 10 \hbar\omega_{\perp}$. For more dilute systems or tighter trapping confinement, the trap can potentially modify pair correlations~\cite{Bruun2014}.

In conclusion, we have presented a novel method to measure in situ particle correlations in quantum gases, and applied it to weakly interacting Bose gases and strongly interacting Fermi gases in two dimensions. Our observation of non-local pairing in the BEC-BCS crossover in the continuum is analogous to that in the case of the attractive Hubbard model~\cite{Hartke2023}.
Important future directions are the study of thermal and quantum fluctuations in these systems, many-particle correlations, spin imbalanced Fermi gases, and microscopic studies on Bose and Fermi polarons. Extensions to Bose-Fermi mixtures, and to spin-resolved imaging for fermions~\cite{Hartke2023} are within close reach. With atom-resolved imaging, one comes close to having complete information about correlations in continuum quantum gases.

We would like to thank Tarik Yefsah for insightful and stimulating discussions. This work was supported by the NSF CUA and PHY-2012110, AFOSR (FA9550-23-1-0402 and MURI), ARO (W911NF-23-1-0382 and DURIP), DOE (DE-SC0024622), DARPA APAQuS and the Vannevar Bush Faculty Fellowship (ONR N00014-19-1-2631). R.J.F. acknowledges funding from the David and Lucile Packard Foundation.

While this work was in progress, we became aware of related work on in situ measurements of correlations in ultracold Fermi~\cite{Yefsah2024} and Bose gases~\cite{Ketterle2024}, as well as a magnifying expansion technique for interacting fermions~\cite{Brandstetter2024}.

\bibliography{microscope.bib}

\begin{thebibliography}{75}
\bibitem [{Pau()}]{Paulihole}%
  \BibitemOpen
  \href@noop {} {}\bibinfo {note} {The term ``Pauli hole'' has been used in the
  past as well, to emphasize the origin in Pauli exclusion.}\BibitemShut
  {Stop}%
\end{thebibliography}

\newpage

\section{Supplemental material}
\setcounter{figure}{0} 
\renewcommand{\thefigure}{S\arabic{figure}} 
\subsection{Preparation of 2D quantum gases}

For the study of bosonic $^{23}$Na, we start with approximately $10^5$ $^{23}$Na atoms prepared in the spin-polarized $|F=1, m_F =1\rangle$ state trapped by the single oblate optical dipole trap of vertical waist $w = 4\,\mu$m. The optical power of the dipole trap is ramped down within $1.2$s, leading to a final anisotropic pancake trap of frequencies $\nu_{x,y,z} = (12(1),17.4(2),360(30))\rm\,Hz$. Subsequently, a vertical magnetic field gradient is turned on to provide further evaporative cooling, resulting in a two-dimensional BEC at the center of the trap. To heat up the gas above the condensation transition temperature, the power of the dipole trap is parametrically modulated at twice the horizontal frequencies. After $360$ms of parametric driving, a wait time of $5$s is added to ensure thermalization of the dilute thermal Bose gas.

For the study of fermionic $^6$Li, we instead start with $10^5$ $^6$Li atoms in an equal mixture of hyperfine states $|F=1/2, m_F=\pm 1/2\rangle$. The evaporation is performed at $800$G, near a broad Feshbach resonance between these two spin states. Following similar steps as in the cooling of $^{23}$Na, the power in the sheet beam is ramped to a final trap with frequencies $\nu_{x,y,z} = (110(7),94(1),3095(35))\rm\,Hz$, and subsequently the magnetic field gradient is applied for further evaporation. This leads to a degenerate two-component Fermi gas containing approximately $40$ atoms in total.

\subsection{Fluctuation Thermometry}

\begin{figure}[h]
 \includegraphics[width=0.8\linewidth]{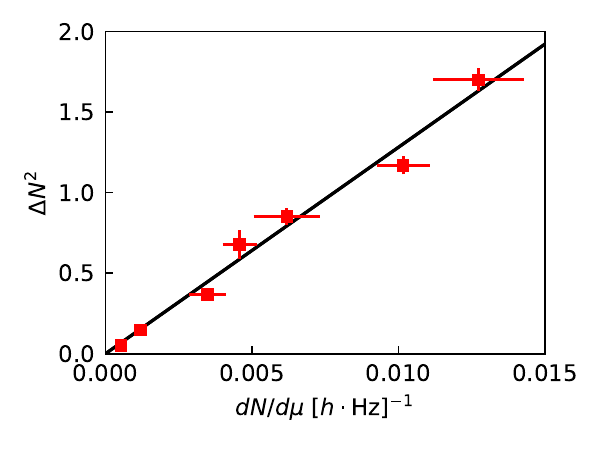}
   \caption{
    Fluctuation thermometry in a weakly interacting Fermi gas. The number fluctuations $\Delta N^2$ in various $9{\times}9$ regions of the pinning lattice are plotted against the local compressibility ${\propto} {\rm d}N/{\rm d}\mu$. The slope yields the temperature $T = 6.1\pm 0.2 \,\rm nK$. The interaction parameter is $\eta{=}\log(k_F a_{\rm 2D}){=}4.2$.
   } 
 \label{fig6:fluctuation}
\end{figure}

In Fig.~\ref{fig6:fluctuation}, we calibrate the temperature of the weakly interacting Fermi gas at $\log (k_Fa_{2D}) = 4.2$ via the Fluctuation Dissipation theorem \cite{Ho2011, Moritz, Koehlthermometry, Hartke2020}: 
\begin{equation}
    \Delta N^2 = k_B T \frac{\partial N}{\partial \mu}
\end{equation}
where $N$, $\Delta N^2$ denote the mean and variance of atom number within a finite volume, and $\mu$ is the local chemical potential. We obtain $\Delta N^2$ and $N$ from a set of local patches of size $9\times 9$ sites over $\sim 100$ experimental realizations. With the calibrated trapping potential, $\frac{\partial N}{\partial \mu} = -\frac{\partial N}{\partial  V}$ for every patch. In Fig.~\ref{fig6:fluctuation} the linear slope of $\Delta N^2$ versus ${\rm d}N/{\rm d}\mu$ yields the temperature of the system. We obtain $T = 6.1(2)$ nK. For fluctuation thermometry to work, the linear patch size must exceed the range of correlations, which in an interacting gas with well-defined sound velocity $c$ is ensured to fall exponentially beyond a distance $r \approx \hbar c/(k_B T)$~\cite{Astrakharchik2007}, which in our case is $\sim 15\,\rm \mu m$. Effectively, already beyond the interparticle spacing $n^{-1/2}= 3.6\,\rm \mu m$ the observed correlations are vanishingly small, and accordingly we found consistent fluctuations when varying the box length beyond this range.

\subsection{Finite Resolution Effects from Lattice Spacing}

The atoms are pinned in an optical lattice of spacing $752$ nm, which thus realizes a discrete spatial grid for the correlation measurement.
As long as correlation functions do not display significant curvature on the scale of the lattice spacing this sampling is expected to give a faithful measurement of correlations. Naturally, the error will be largest at short range, where two-particle correlations e.g. from fermion pairing are strong. To test the method, we numerically simulate the effect using the mean-field form of correlation functions for various interaction parameters $\eta = \log(k_Fa_{2D})$ explored in the text. We bin the ideal two-dimensional correlation map with sample size $752$ nm, then compare the pair correlation $g^{(2)}(r)$ obtained on the discretized distances after binning to the original curves. We find that even for the distance of one lattice site, the values agree within $5\%$ for all parameters we have explored.

 \begin{figure*}[!htbp]
 \includegraphics[width=0.8\linewidth]{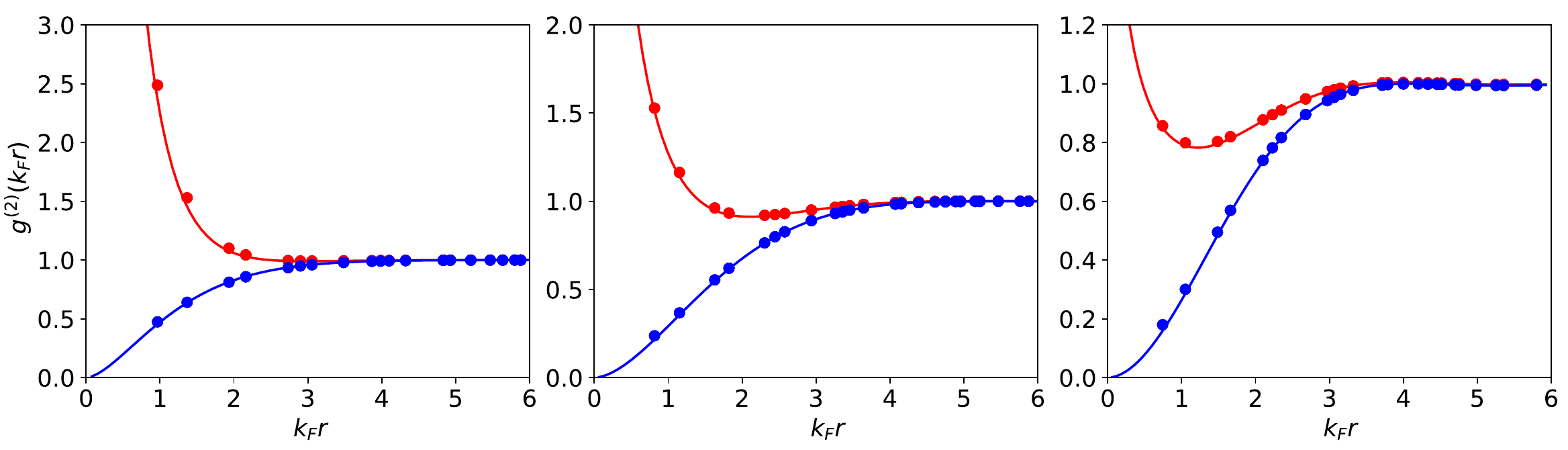}
   \caption{
        Pair correlations before and after sampling with a finite lattice spacing of $752$ nm. From left to right $\eta=\log(k_Fa_{2D})=0.2$, $1.2$ and $4.2$. The solid curves are the mean-field correlation functions Eq.~\ref{eq:correlations} with parameters $c$ and $b$ obtained by fitting to the experimental data. Dots are correlations after binning with the lattice grid of spacing $752$ nm. 
        } 
 \label{figSI:correlation}
\end{figure*}

\end{document}